\documentclass[sn-mathphys-num]{sn-jnl}

\usepackage{array}
\usepackage{graphicx}%
\usepackage{subfigure}
\usepackage{multirow}%
\usepackage{amsmath,amssymb,amsfonts}%
\usepackage{amsthm}%
\usepackage{mathrsfs}%
\usepackage[title]{appendix}%
\usepackage{xcolor}%
\usepackage{textcomp}%
\usepackage{manyfoot}%
\usepackage{booktabs}%
\usepackage{tabularx}
\usepackage{algorithm}%
\usepackage{algorithmicx}%
\usepackage{algpseudocode}%
\usepackage{listings}%
\usepackage{float}

\usepackage{graphicx}
\usepackage{array}
\usepackage{multirow}
\usepackage{booktabs}
\usepackage{hyperref}

\usepackage{longtable}
\usepackage{url}
\newcommand{\tabincell}[2]{\begin{tabular}{@{}#1@{}}#2\end{tabular}}

\theoremstyle{thmstyleone}%
\theoremstyle{thmstyletwo}%

\theoremstyle{thmstylethree}%

\begin{document}

\title[A Survey on Large Language Models for Recommendation]{A Survey on Large Language Models for Recommendation}


\author[1,2]{Likang Wu}\email{wulk@mail.ustc.edu.cn}

\author[1,2]{Zhi Zheng}\email{zhengzhi97@mail.ustc.edu.cn}
\equalcont{Equal contribution.}

\author[2]{Zhaopeng~Qiu}\email{zhpengqiu@gmail.com}
\equalcont{Equal contribution.}

\author*[1]{Hao Wang}\email{wanghao3@ustc.edu.cn}

\author[1]{Hongchao Gu}\email{hcgu@mail.ustc.edu.cn}

\author[1]{Tingjia~Shen}\email{jts\_stj@mail.ustc.edu.cn}

\author[2]{Chuan Qin}\email{chuanqin0426@gmail.com}

\author[2]{Chen Zhu}\email{zc3930155@gmail.com}

\author*[2]{Hengshu Zhu}\email{zhuhengshu@gmail.com}

\author[1]{Qi Liu}\email{qiliuql@ustc.edu.cn}

\author*[3]{Hui Xiong}\email{xionghui@ust.hk}

\author*[1]{Enhong Chen}\email{cheneh@ustc.edu.cn}

\affil[1]{\orgdiv{State Key
Laboratory of Cognitive Intelligence}, \orgname{University of Science and Technology of China}, \orgaddress{\street{JinZhai Road}, \city{Hefei}, \postcode{230026}, \state{Anhui}, \country{China}}}

\affil[2]{\orgdiv{Career Science Lab}, \orgname{BOSS Zhipin}, \city{Beijing}, \postcode{100020}, \country{China}}

\affil[3]{\orgname{Hong Kong University of Science and Technology (Guangzhou)}, \city{Guangzhou}, \postcode{510000}, \state{Guangdong}, \country{China}}


\abstract{Large Language Models (LLMs) have emerged as powerful tools in the field of Natural Language Processing (NLP) and have recently gained significant attention in the domain of Recommendation Systems (RS). These models, trained on massive amounts of data using self-supervised learning, have demonstrated remarkable success in learning universal representations and have the potential to enhance various aspects of recommendation systems by some effective transfer techniques such as fine-tuning, prompt tuning, etc. The crucial aspect of harnessing the power of language models in enhancing recommendation quality is the utilization of their high-quality representations of textual features and their extensive coverage of external knowledge to establish correlations between items and users. To provide a comprehensive understanding of the existing LLM-based recommendation systems, this survey presents a taxonomy that categorizes these models into two major paradigms, respectively Discriminative LLM for Recommendation (DLLM4Rec) and Generative LLM for Recommendation (GLLM4Rec), with the latter being systematically sorted out for the first time. Furthermore, we systematically review and analyze existing LLM-based recommendation systems within each paradigm, providing insights into their methodologies, techniques, and performance. Additionally, we identify key challenges and several valuable findings to provide researchers and practitioners with inspiration. We have also created a GitHub repository to index relevant papers and resources on LLMs for recommendation\footnote{https://github.com/WLiK/LLM4Rec-Awesome-Papers}. }

\keywords{Large Language Models, Recommendation System}



\maketitle

\section{Introduction}\label{sec1}
Recommendation systems play a critical role in assisting users in finding relevant and personalized items or content. With the emergence of Large Language Models (LLMs) in Natural Language Processing (NLP), there has been a growing interest in harnessing the power of these models to enhance recommendation systems. 

The key advantage of incorporating LLMs into recommendation systems lies in their ability to extract high-quality representations of textual features and leverage the extensive external knowledge encoded within them~\cite{liu2023pre}. And this survey views LLM as the Transformer-based model with a large number of parameters, trained on massive datasets using self/semi-supervised learning techniques, e.g., BERT, GPT series, PaLM series, etc\footnote{https://en.wikipedia.org/wiki/Large\_language\_model}. Unlike traditional recommendation systems, the LLM-based models capture contextual information, and comprehend user queries, item descriptions, and other textual data more effectively~\cite{DBLP:conf/recsys/Geng0FGZ22}. By understanding the context, LLM-based RS can improve the accuracy and relevance of recommendations, leading to enhanced user satisfaction. Meanwhile, facing the common data sparsity issue of limited historical interactions~\cite{da2020recommendation}, LLMs also bring new possibilities to recommendation systems through zero/few-shot recommendation capabilities~\cite{DBLP:conf/ecir/SileoVR22}. These models can generalize to unseen candidates due to the extensive pre-training with factual information, domain expertise, and common-sense reasoning, enabling them to provide reasonable recommendations even without prior exposure to specific items or users. 

\begin{figure}[t]
    \centering
    \includegraphics[width=0.7\textwidth]{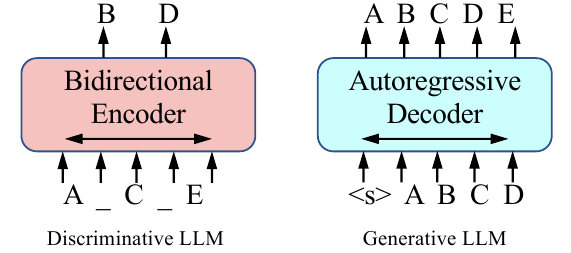}\
    \caption{Two major training paradigms of large language models: Discriminative LLM (e.g. BERT) and Generative LLM (e.g. GPT).}
    \label{fig:DLLM}
\end{figure}

The aforementioned strategies are already well-applied in discriminative models. However, with the evolution of AI learning paradigms, generative language models have started to gain prominence~\cite{zhao2023survey} as shown in Fig.~\ref{fig:DLLM}. A prime example of this is the emergence of ChatGPT and other comparable models, which have significantly disrupted human life and work patterns. Furthermore, the fusion of generative models with recommendation systems offers the potential for even more innovative and practical applications. For instance, the interpretability of recommendations can be improved, as LLM-based systems are able to provide explanations based on their language generation capabilities~\cite{DBLP:journals/corr/abs-2303-14524}, helping users understand the factors influencing the recommendations. Moreover, generative language models enable more personalized and context-aware recommendations, such as users' customizable prompts~\cite{li2023personalized} in the chat-based recommendation system, enhancing user engagement and satisfaction with the diversity of results.

\begin{figure*}[t]
    \centering
    \includegraphics[width=1.\textwidth]{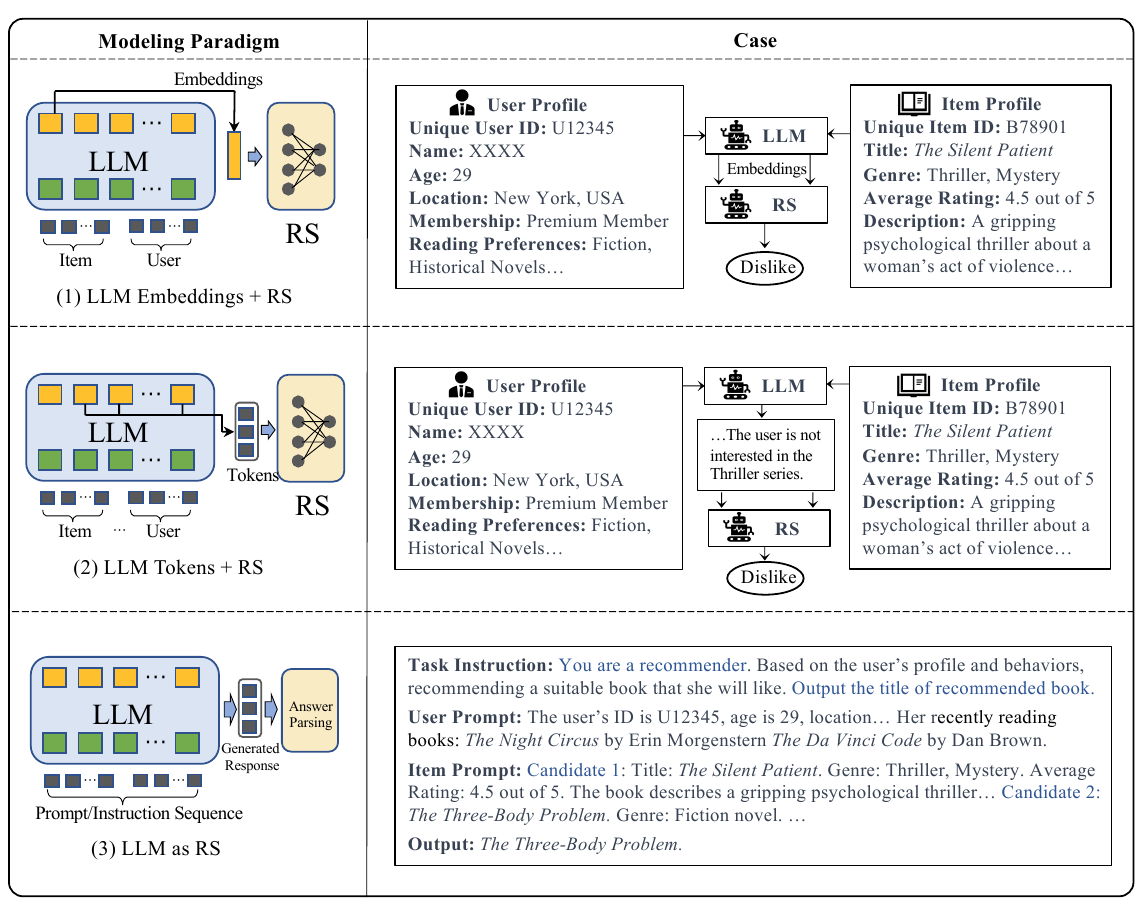}
    \caption{Three representative modeling paradigms of the research for large language models on recommendation systems.}
    \label{fig:modeling}
\end{figure*}
Motivated by the remarkable effectiveness of the aforementioned paradigms in solving data sparsity and efficiency issues, the adaptation of language modeling paradigms for recommendation has emerged as a promising direction in both academia and industry, significantly advancing the state-of-the-art in the research of recommendation systems. 
In the early stage, there are a few studies that review relevant papers in this domain~\cite{zeng2021knowledge,liu2023pre}. ~\cite{zeng2021knowledge} summarizes some research on the pre-training of recommendation models and discusses knowledge transfer methods between different domains. ~\cite{liu2023pre} proposes an orthogonal taxonomy to divide existing pre-trained language model-based recommendation systems w.r.t. their training strategies and objectives, analyzes and summarizes the connection between pre-trained language model-based training paradigms and different input data types. However, both of these surveys primarily focus on the transfer of training techniques and strategies in pretraining language models, rather than exploring the potential of language models and their capabilities, i.e., LLM-based way. Additionally, they lack a comprehensive overview of the recent advancements and systematic introductions of generative large language models in the recommendation field. To address this issue, we delve into LLM-based recommendation systems, categorizing them into discriminative LLMs for recommendation and generative LLMs for recommendation, and the focus of our review is on the latter. Recently, there have been several reviews introducing the application of large language models in recommendation systems or related technologies~\cite{lin2024recommender,zhao2024recommender,li2024large,chen2023large}. However, our paper is the first to comprehensively summarize three representative modeling paradigms of applying large language models in recommendation systems. Its distinct characteristics of being concise yet broadly and accurately covered have garnered significant attention in the industry. Additionally, our paper thoroughly reviews the major challenges faced by large language models in the recommendation field, providing valuable insights to guide future research directions in this area. The main contributions of our survey are summarized as follows: 

\begin{itemize}
\item We present a systematic survey of the current state of LLM-based recommendation systems, focusing on expanding the capacity of language models. We provide a systematic overview of related advancements and applications by analyzing the existing methods.
\item From the perspective of modeling paradigms, we categorize the current studies of large language model recommendations into three distinct schools of thought. Any existing method can be fittingly placed within these categories, thereby providing a clear and organized overview of this burgeoning field.
\item Our survey critically analyzes the advantages, disadvantages, and limitations of existing methods. We identify key challenges faced by LLM-based recommendation systems and propose valuable findings that can inspire further research in this potential field. 
\end{itemize}

\section{Modeling Paradigms and Taxonomy}
The basic framework of all large language models is composed of several transformer blocks, e.g., GPT, PaLM, LLaMA, etc. The input of this architecture is generally composed of token embeddings or position embeddings and so on, while the expected output embedding or tokens can be obtained at the output module. Here, both the input and output data types are textual sequences. As shown in (1)-(3) in Figure~\ref{fig:modeling}, for the adaption of language models in recommendations, i.e., the modeling paradigm, existing work can be roughly divided into the following three categories:
\begin{enumerate}[(1)]
    \item \textbf{LLM Embeddings + RS}. This modeling paradigm views the language model as a feature extractor, which feeds the features of items and users into LLMs and outputs corresponding embeddings. A traditional RS model can utilize knowledge-aware embeddings for various recommendation tasks.
    \item \textbf{LLM Tokens + RS}. Similar to the former method, this approach generates tokens based on the inputted items' and users' features. The generated tokens capture potential preferences through semantic mining, which can be integrated into the decision-making process of a recommendation system.
    \item \textbf{LLM as RS}. Different from (1) and (2), this paradigm aims to directly transfer pre-trained LLM into a powerful recommendation system. The input sequence usually consists of the profile description, behavior prompt, and task instruction. The output sequence is expected to offer a reasonable recommendation result.
\end{enumerate}

\begin{figure}[t]
    \centering
    \includegraphics[width=0.85\textwidth]{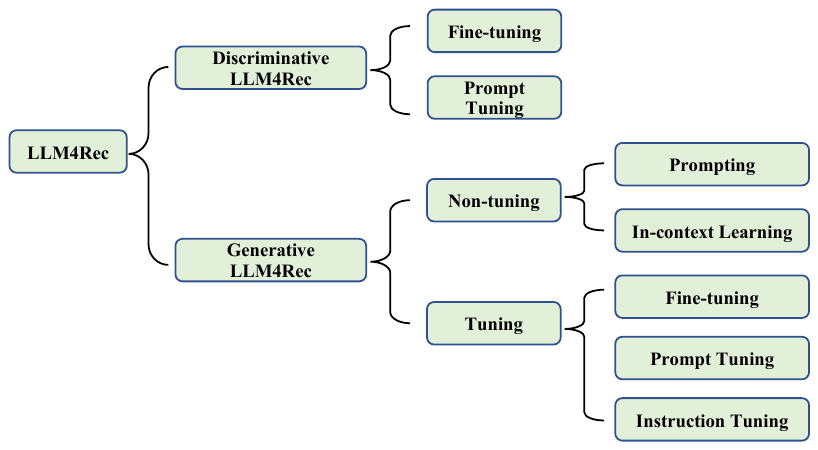}
    \caption{A taxonomy of the research for large language models on recommendation systems.}
    \label{fig:taxonomy}
\end{figure}
\begin{figure}[t]
    \centering
    \includegraphics[width=0.88\textwidth]{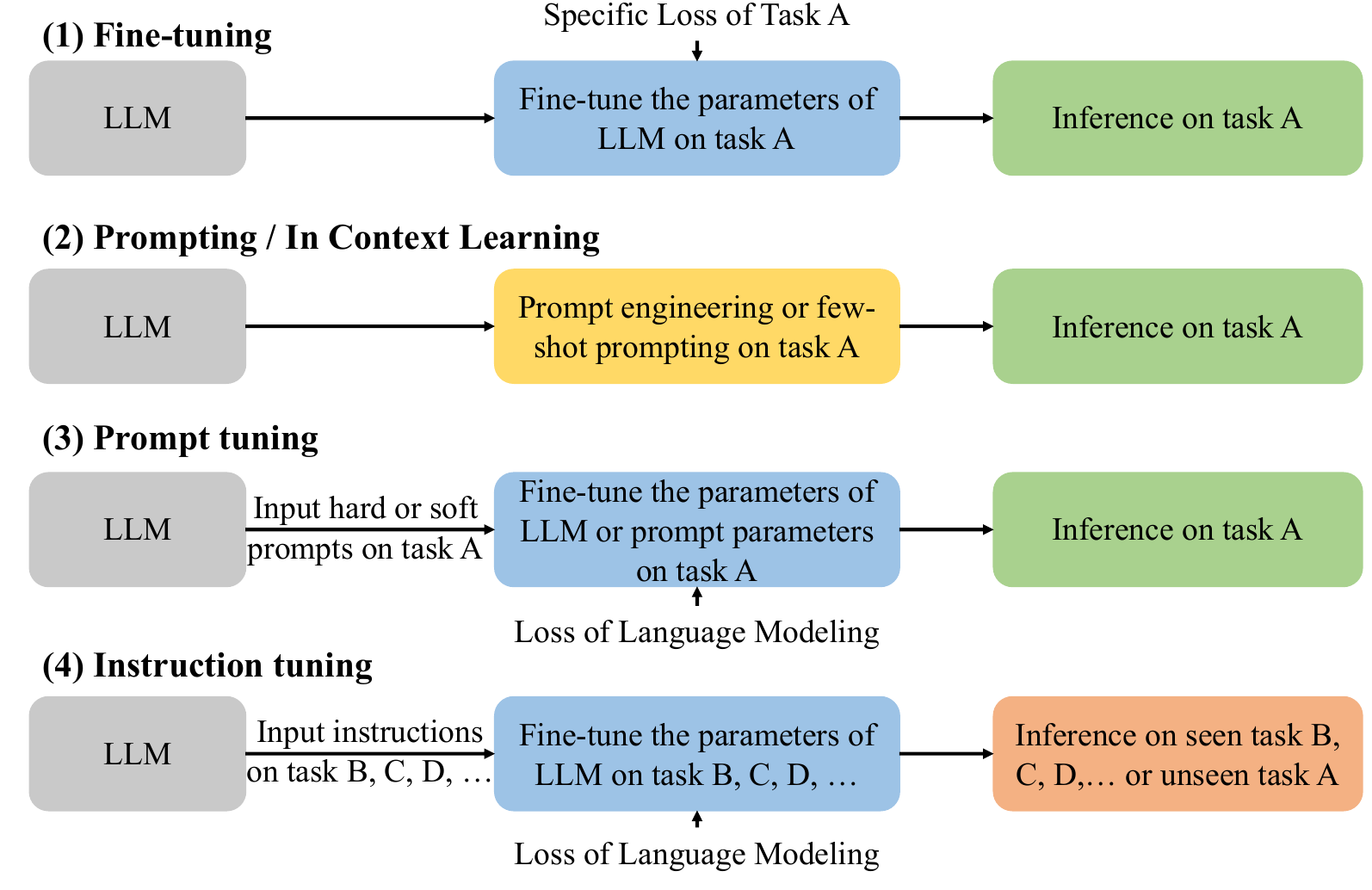}
    \caption{Detailed explanation of five different training (domain adaption) manners for LLM-based recommendations.}
    \label{fig:training}
\end{figure}

In practical applications, the choice of large language models significantly influences the design of modeling paradigms in recommendation systems. As shown in Figure~\ref{fig:taxonomy}, in this paper, we categorize existing works into two main categories, discriminative LLMs and generative LLMs for recommendation, respectively. The taxonomy of development styles of LLMs for recommendation can be further subdivided based on the training manner, and the distinction among different manners is illustrated clearly in Figure~\ref{fig:training}. Generally, discriminative language models are well-suited for embedding within the paradigm (1), while the response generation capability of generative large language models further supports paradigms (2) or (3). 

\section{Discriminative LLMs for Recommendation}
Indeed, so-called discriminative large language models in the recommendation area mainly refer to those models of BERT series~\cite{devlin2018bert}. Due to the expertise of discriminative language models in natural language understanding tasks, they are often considered as embedding backbones for different downstream tasks. This holds true for recommendation systems as well. Most existing works align the representations of pre-trained models like BERT with the domain-specific data through fine-tuning. Additionally, some research explores training strategies like prompt tuning and adapter tuning. The representative approaches and common-used datasets are listed in Table~\ref{tab:list} and Table~\ref{tab:data}.
\subsection{Fine-tuning}
\label{sec:Discriminative Fine-tuning}
\begin{figure*}[t]
  \centering
  \subfigure[Finetuning DLLM for recommendation]{
    \includegraphics[width=0.485\columnwidth]{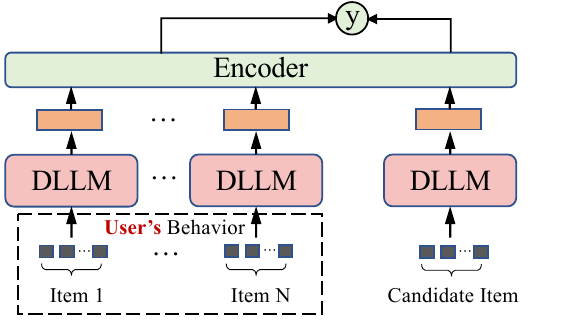}}
  \subfigure[Prompt tuning DLLM for recommendation]{
    \includegraphics[width=0.485\columnwidth]{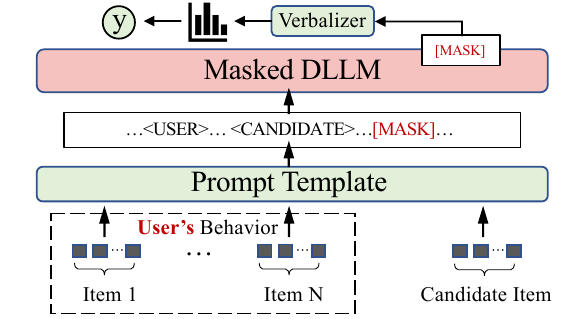}}
  \caption{Discriminative LLMs for recommendation. }
  \label{fig:DLLM4rec} 
\end{figure*}

Fine-tuning pre-trained language models is a universal technique that has gained significant attention in various natural language processing (NLP) tasks, including recommendation systems. The idea behind fine-tuning is to take a language model, which has already learned rich linguistic representations from large-scale text data, and adapt it to a specific task or domain by further training it on task-specific data. The classic architect is shown in Fig~\ref{fig:DLLM4rec} (a).

The process of fine-tuning involves initializing the pre-trained language model with its learned parameters and then training it on a recommendation-specific dataset. This dataset typically includes user-item interactions, textual descriptions of items, user profiles, and other relevant contextual information. During fine-tuning, the model's parameters are updated based on the task-specific data, allowing it to adapt and specialize for target recommendation tasks. The learning objectives in the pre-training and fine-tuning stages can vary, as they are aimed at different optimization targets.

Since the fine-tuning strategy is flexible, most BERT-enhanced recommendation methods can be summarized into this track.  For the basic representation task,~\cite{qiu2021u} proposed a novel pre-training and fine-tuning-based approach U-BERT to learn users' representation, which leveraged content-rich domains to complement those users' feature with insufficient behavior data. A review co-matching layer is designed to capture implicit semantic interactions between the reviews of users
and items. Similarly, in UserBERT~\cite{wu2021userbert}, two
self-supervision tasks are incorporated for user model pre-training on unlabeled behavior data to empower user modeling. This model utilizes medium-hard contrastive learning, masked behavior prediction, and behavior sequence matching to train accurate user representation via captured inherent user interests and relatedness.

The pre-trained BERT achieved outstanding breakthroughs in the ranking task as well. BECR~\cite{yang2022lightweight} proposed a lightweight composite re-ranking scheme that combined deep contextual token interactions and traditional lexical term-matching features at the same time. With a novel composite token encoding, BECR effectively approximates the query representations using pre-computable token embeddings based on
uni-grams and skip-n-grams, allowing for a reasonable tradeoff between ad-hoc ranking relevance and efficiency. Besides, ~\cite{wu2022multi} proposed an end-to-end multi-task learning framework for product ranking with fine-tuned domain-specific BERT to address the issue of vocabulary mismatch between queries and products. The authors utilized the mixture-of-experts layer and probability transfer between tasks to harness the abundant engagement data. In a more specific situation of code example recommendations, the authors revealed that utilizing natural language queries (BERT + LSH) yields better ranking results compared to API-based queries~\cite{rahmani2023improving}.

There are also many related studies in other specific tasks or scenarios, e.g., group recommendation~\cite{zhang2022gbert}, search/matching~\cite{yao2022reprbert}, CTR prediction~\cite{muhamed2021ctr}. Especially, the ``pre-train, fine-tuning'' mechanism played an important role in several sequential or session-based recommendation systems, such as BERT4Rec~\cite{Sun:2019:BSR:3357384.3357895}, RESETBERT4Rec~\cite{zhao2022resetbert4rec}, and Adapter Tuning~\cite{fu2023exploring,hu2024enhancing}. However, the above models only leveraged the advantages of the training strategy rather than expanding the large language model into the recommendation field, so it was not the focus of our discussion. 
The sequence representation learning model UniSRec ~\cite{hou2022towards} developed a BERT-fine-tuned framework, which associated description text of items to learn transferable representations across different recommendation scenarios. Considering the binding between item text and item representations might be too tight, leading to potential problems such as over-emphasizing the effect of text features and exaggerating the negative impact of domain gap, Hou et al~\cite{houlearningv} proposed to learn distinguishable vector-quantized item codes for transferable sequential recommenders.
For the content-based recommendation, especially news recommendation, NRMS~\cite{wu2021empowering}, Tiny-NewsRec~\cite{yu2022tiny}, PREC~\cite{liu2022boosting}, exploited large language models to empower news recommendation via handling known domain shift problems or reducing transfer cost. Specifically, to answer the crucial question that \emph{Can a purely modality-based recommendation model (MoRec) outperforms or matches a pure ID-based model (IDRec) by replacing the itemID embedding with a SOTA modality encoder?}, \cite{yuan2023go} conducted large-scale experiments and found that modern MoRec could already perform on par or better than IDRec with the typical recommendation architecture (i.e., Transformer backbone) even in the non-cold-start item recommendation setting with the SOTA and E2E-trained Modality Encoder. The subsequent exploration~\cite{li2023exploring} based on larger-scale language model encoders, e.g. OPT~\cite{zhang2022opt}, further validated the viewpoint.

In summary, the integration of BERT fine-tuning into recommendation systems fuses the powerful external knowledge and personalized user preference, which primarily aims to promote recommendation accuracy and simultaneously obtains a little cold-start handling capability for new items with limited historical data.
\subsection{Prompt Tuning}
Instead of adapting LLMs to different downstream recommendation tasks by designing specific objective functions, prompt tuning~\cite{lester2021power} aims to align the objective of recommendation tuning with pre-trained loss, using hard/soft prompts and a label word verbalizer. As shown in Fig~\ref{fig:DLLM4rec} (b), due to the mask-based training commonly employed in DLLM, the role of the mentioned verbalizer is to establish a mapping between the words predicted by DLLM at the [MASK] position and the actual labels. This association allows for the linkage between the language model and the task, ensuring their alignment.
For example,~\cite{penha2020does} leveraged BERT's Masked Language Modeling (MLM) head to uncover its understanding of item genres using cloze-style prompts. They further utilized BERT's Next Sentence Prediction (NSP) head and similarity (SIM) of representations to compare relevant and non-relevant search and recommendation query-document inputs. The experiment showed that BERT, even without any fine-tuning, can prioritize relevant items in the ranking process.~\cite{yang2022improving} developed a conversational recommendation system with prompts, where a BERT-based item encoder directly mapped the metadata of each item to an embedding. Similarly, Shen et al~\cite{shen2023towards} developed a conversational recommendation system that incorporates user-item attribute fairness analysis. They achieved this by employing constructed prompt templates with placeholders (referred to as template-based result generation). These templates include non-preferential information such as names or relationships, which can implicitly indicate characteristics like race, gender, sexual orientation, geographical context, and religion. The analysis demonstrated that by combining train side masking and test side neutralization of non-preferential entities, the observed biases can be eliminated without compromising recommendation performance.  Recently, Prompt4NR~\cite{zhang2023prompt} pioneered the application of the prompt learning paradigm for news recommendation. This framework redefined the objective of predicting user clicks on candidate news as a cloze-style mask-prediction task.
The experiments found that the performance of recommendation systems is noticeably enhanced through the utilization of multi-prompt ensembling, surpassing the results achieved with a single prompt on discrete and continuous templates. This highlights the effectiveness of prompt ensembling in combining multiple prompts to make more informed decisions.

\section{Generative LLMs for Recommendation}
Compared to discriminative models, generative models have better natural language generation capabilities.
Therefore, unlike most discriminative model-based approaches that align the representation learned by LLMs to the recommendation domain, most generative model-based work translates recommendation tasks as natural language tasks, and then applies techniques such as in-context learning, prompt tuning, and instruction tuning to adapt LLMs to directly generate the recommendation results.
Moreover, with the impressive capabilities demonstrated by ChatGPT, this type of work has received increased attention recently.

As shown in Figure~\ref{fig:taxonomy}, according to whether tuning parameters, these generative LLM-based approaches can be further subdivided into two paradigms: \textit{non-tuning paradigm} and \textit{tuning paradigm}. \textbf{Here the tuning/non-tuning target denotes the used LLM module in the following methods.}
The following two sub-sections will address their details, respectively. The representative approaches and common-used datasets are also listed in Table~\ref{tab:list} and Table~\ref{tab:data}.

\subsection{Non-tuning Paradigm}
The LLMs have shown strong zero/few-shot abilities in many unseen tasks~\cite{GPT3,InstructGPT}. Hence, some recent works assume LLMs already have the recommendation abilities, and attempt to trigger these abilities by introducing specific prompts. 
They employ the recent practice of Instruction and In-Context Learning~\cite{GPT3} to adopt the LLMs to recommendation tasks without tuning model parameters. 
According to whether the prompt includes the demonstration examples, the studies in this paradigm mainly belong to the following two categories: \textit{prompting} and \textit{in-context learning}. 
\subsubsection{Prompting}





This category of work aims to design more suitable instructions and prompts to help LLMs better understand and solve the recommendation tasks. 
~\cite{DBLP:journals/corr/abs-2304-10149} systematically evaluated the performance of ChatGPT on five common recommendation tasks, i.e., \textit{rating prediction}, \textit{sequential recommendation}, \textit{direct recommendation}, \textit{explanation generation}, and \textit{review summarization}. They proposed a general recommendation prompt construction framework, which consists of the following elements: (1) task description, adapting recommendation tasks to natural language processing tasks; (2) behavior injection, incorporating user-item interaction to aid LLMs in capturing user preferences and needs; (3) format indicator, constraining the output format and making the recommendation results more comprehensible and assessable. Similarly, ~\cite{DBLP:journals/corr/abs-2305-02182} conducted an empirical analysis of ChatGPT's recommendation abilities on three common information retrieval tasks, including point-wise, pair-wise, and list-wise ranking.
They proposed different prompts for different kinds of tasks and introduced the \textit{role} instructions (such as \textit{You are a news recommendation system now.}) at the beginning of the prompts to enhance the domain adaption ability of ChatGPT.
~\cite{DBLP:journals/corr/abs-2305-04518} explored the feasibility of developing an Artificial General Recommender (AGR) using Large Language Models (LLMs) from the perspective of ten fundamental principles, such as \textit{contextual memory}, \textit{repair mechanism} and \textit{feedback mechanism}.

To evaluate the enhancement of different prompting inputs, ~\cite{sanner2023large} designed three prompt templates for the case of Items only (the attribute of items), Language only (the description of user's preference), and combined Language+Items in their experiments. After analyzing the performance of language models, they discovered that zero-shot and few-shot strategies are highly effective for making recommendations based solely on language-based preferences (without considering item preferences). 
In fact, these strategies have proven to be remarkably competitive in comparison to item-based collaborative filtering methods, particularly in near cold-start scenarios. Meanwhile, to summarize the user's intention by prompt based on their interaction data, MINT~\cite{mysore2023large} employed InstructGPT, a 175B parameter LLM, to generate a synthetic narrative query. This query was then filtered using a smaller language model, and retrieval models were trained on both the synthetic queries and user items. The results indicate that the resulting models outperformed several strong baseline models and ablated models. In a one-shot setup, these models matched or even outperformed a 175B LLM that was directly used for narrative-driven recommendation. However, these methods have not considered decomposing the topics in a textual description, which would result in noisy and target-unclear prompts. KAR~\cite{Xi2023TowardsOR} solved this issue by introducing factorization prompting to elicit accurate reasoning on user preferences and factual knowledge. 

Instead of proposing a general framework, some works focus on designing effective prompts for specific recommendation tasks.
~\cite{DBLP:conf/ecir/SileoVR22} mined the movie recommendation prompts from the pre-training corpus of GPT-2.
~\cite{DBLP:journals/corr/abs-2305-08845} introduced two prompting methods to improve the sequential recommendation ability of LLMs: \textit{recency-focused sequential prompting}, enabling LLMs to perceive the sequential information in the user interaction history, and \textit{bootstrapping}, shuffling the candidate item list multiple times and taking the average scores for ranking to alleviate the position bias problem. 
Due to the limited number of input tokens allowed for the LLMs, it is hard to input a long candidate list in the prompt.
To solve this problem, ~\cite{DBLP:journals/corr/abs-2304-09542} proposed a sliding window prompt strategy, which only ranks the candidates in the window each time, then slides the window in back-to-first order, and finally repeats this process multiple times to obtain the overall ranking results.
~\cite{DBLP:journals/corr/abs-2310-20487} designed a sequence-residual prompt to use LLMs to improve the interpretability of traditional sequential recommender.
~\cite{DBLP:journals/corr/abs-2404-11960} proposed a multi-perspective criteria ensemble framework that improves the consistency and comprehensiveness of pointwise LLM rankers by simulating a virtual annotation team with diverse expertise.
For conversational recommendation tasks, ~\cite{DBLP:conf/cikm/HeXJSLFMKM23} provided empirical evidence that LLMs can outperform specialized models without the need for fine-tuning. Additionally, the authors constructed a new real-world dataset by extracting conversations from the popular website Reddit.

In addition to taking LLMs as recommendation systems, some studies also utilize LLMs to construct model features to improve the conventional recommender system.
~\cite{DBLP:conf/recsys/AcharyaSO23,DBLP:journals/corr/abs-2403-03424,DBLP:journals/corr/abs-2403-18348} and LLM-Rec~\cite{DBLP:journals/corr/abs-2307-15780} used LLMs and prompting strategies to conduct content augmentation to enhance the features of item perspective.
From the user feature perspective, NIR~\cite{DBLP:journals/corr/abs-2304-03153} designed prompts to generate user preference description and LLMRG~\cite{DBLP:journals/corr/abs-2308-10835} used ChatGPT and knowledge base to construct reasoning graph to enhance user representations.
GENRE~\cite{DBLP:journals/corr/abs-2305-06566} introduced three prompts to employ LLMs to conduct three feature enhancement sub-tasks for news recommendation from both user and item perspectives.
Specifically, it used ChatGPT to refine the news titles according to the abstract, extract profile keywords from the user reading history, and generate synthetic news to enrich user historical interactions.
Similarly, LLMRec~\cite{DBLP:journals/corr/abs-2311-00423} and RLMRec~\cite{DBLP:journals/corr/abs-2310-15950} first used ChatGPT to generate textual features for users and items and then took these features to enhance the ID-based representation learning.

In practice, in addition to the ranking model, the whole recommendation system generally consists of multiple important components, such as a content database and a candidate retrieval model.
Hence, another line of using LLMs for recommendation is taking them as the controllers of the whole system.
ChatREC~\cite{DBLP:journals/corr/abs-2303-14524}, RAH~\cite{DBLP:journals/corr/abs-2308-09904}, BiLLP~\cite{Shi2024LargeLM} and InteRecAgent~\cite{DBLP:journals/corr/abs-2308-16505} designed the interactive recommendation framework around LLMs, which understands user requirements through multi-turn dialogues, and calls existing recommendation systems and various tools, such as database, retriever, memory, to provide results. The agent can play a crucial role in such conversational situations, thus several advanced versions are developed to optimize the fusion between chat intelligence and recommender module~\cite{jin2024lending,huang2024can}. 
What's more, some recent agent-based models~\cite{DBLP:journals/corr/abs-2402-18240,zhang2024agentcf} propose recommendation frameworks for LLM-based agent platforms, emphasizing personalized agent services through enhanced interaction and collaboration among users, agent recommenders, and agent items.
GeneRec~\cite{DBLP:journals/corr/abs-2304-03516} proposed a generative recommendation framework and used LLMs to control when to recommend existing items or to generate new items by AIGC models. 
Furthermore, ~\cite{DBLP:journals/corr/abs-2403-09738}, RecAgent~\cite{wang2023recagent} and Agent4Rec~\cite{DBLP:journals/corr/abs-2310-10108} further utilized LLM as intelligent simulator to develop a virtual recommendation environment. The simulator typically consists of two main modules: user and recommender. The user module enables browsing the recommendation site, interaction with other users, and posting on social media. The recommender module offers tailored search and recommendation lists, supporting various model designs for recommendation. Users in the environment interact based on LLM-generated responses, evolving organically to mirror real-world behavior. 
These projects show potential utilization across several applications, such as simulating the feedback for RL-based recommendations, tracking information dissemination process among the users on social media, investigating the filter bubble effect, and unveiling the underlying causal relationships embedded within recommender system scenarios.
Instead of using LLMs as controllers, UniLLMRec~\cite{DBLP:journals/corr/abs-2404-00702} proposed an end-to-end chain-like recommendation framework that leverages LLMs to efficiently integrate recall, ranking, and re-ranking tasks.

In summary, these studies employ natural language prompts to leverage the zero-shot capabilities of LLMs for recommendation tasks, offering a cost-effective and pragmatic approach.

\begin{figure*}[t]
    \centering
    \includegraphics[width=1.\textwidth]{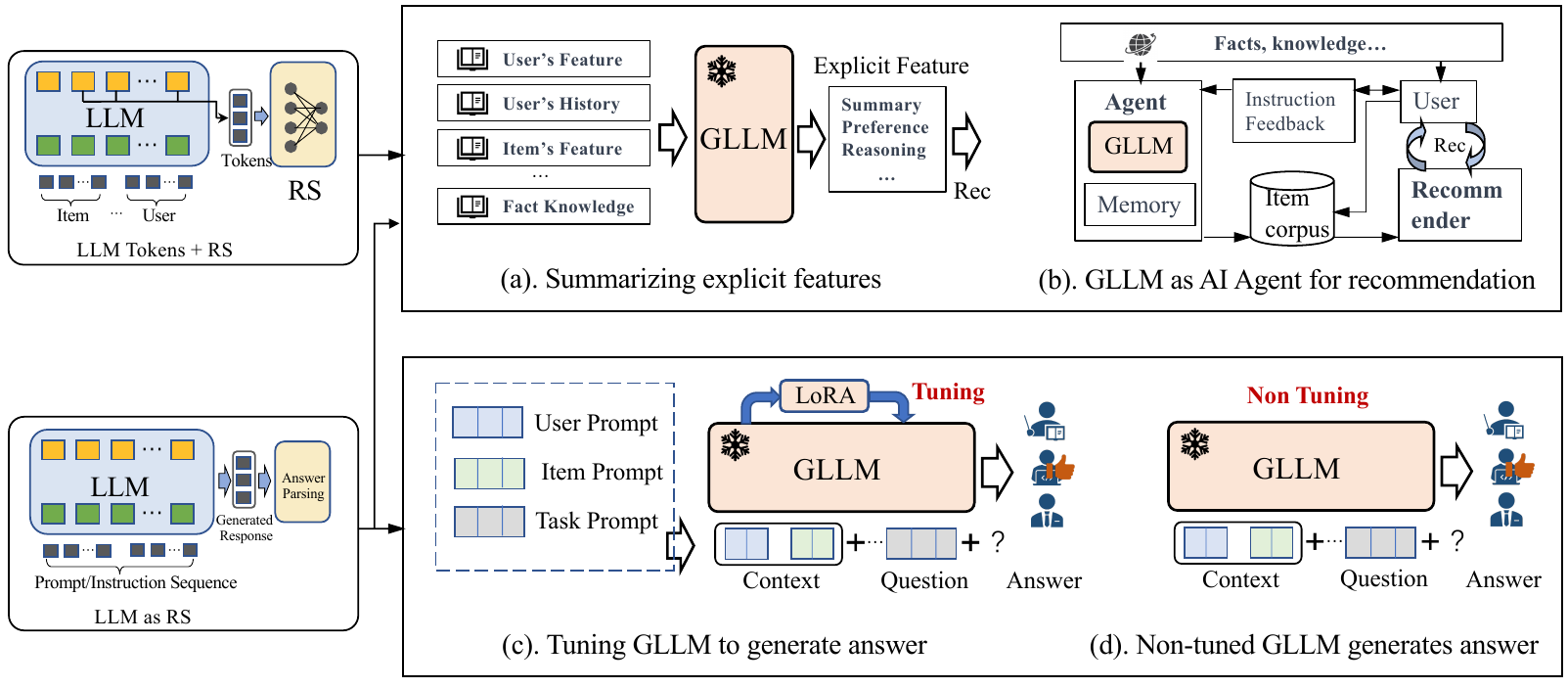}
    \caption{Generative LLMs for recommendation.}
    \label{fig:GLLM4rec}
\end{figure*}
\subsubsection{In-context Learning}
In-context learning is a technique used by GPT-3 and other LLMs to quickly adapt to new tasks and information. 
With a few demonstration input-label pairs, they can predict the label for an unseen input without additional parameter updates~\cite{DBLP:journals/corr/abs-2212-10559}.
Hence, some works attempt to add demonstration examples in the prompt to make LLMs better understand the recommendation tasks.
For sequential recommendation, ~\cite{DBLP:journals/corr/abs-2305-08845} introduced demonstration examples by augmenting the input interaction sequence itself. In detail, they paired the prefix of the input interaction sequence and the corresponding successor as examples.
~\cite{DBLP:journals/corr/abs-2403-10135} investigated the
effects of instruction format, task consistency,
demonstration selection, and number of demonstrations
~\cite{DBLP:journals/corr/abs-2304-10149} and ~\cite{DBLP:journals/corr/abs-2305-02182} designed the demonstration example templates for various recommendation tasks and the experimental results also showed the in-context learning method will improve the recommendation abilities of LLMs on most tasks. In addition, a suitable demonstration can be used to control the output format and content of the LLM~\cite{wang2023rethinking}, which can improve the regular evaluation metric. This is crucial for developing a stable and robust recommender system.

However, in comparison to prompting, only a few studies have explored the use of In-context Learning of Language Models (LLMs) in recommendation tasks. Numerous open questions remain, including the selection of demonstration examples and the influence of the number of demonstration examples on recommendation performance.

\subsection{Tuning Paradigm}
As we mentioned above, LLMs have strong zero/few-shot abilities, and their recommendation performance can significantly surpass random guessing with appropriate prompt design. However,  it is not surprising that recommendation systems constructed in this manner fail to surpass the performance of recommendation models trained specifically for a given task on specific data. Therefore, many researchers aim to enhance the recommendation ability of LLMs by further fine-tuning or prompt learning. In this paper, we categorize the paradigm of the tuning methods into three different types, respectively fine-tuning, prompt tuning, and instruction tuning. Specifically, in the fine-tuning paradigm, the usage methods for discriminative and generative large language models are notably similar. The LLMs mainly serve as encoders to extract representations of users or items, and the parameters of the LLMs are subsequently fine-tuned on the specific loss functions of downstream recommendation tasks. Meanwhile, in the prompt tuning and instruction tuning paradigms, the output of the large models is consistently textual, and their parameters are trained using the loss of language modeling. The primary distinction between the prompt tuning and instruction tuning training paradigms is that prompt tuning predominantly focuses on a specific task, e.g., rating prediction, while the LLMs are trained for multiple tasks with different types of instructions under the instruction tuning paradigm. Therefore, the LLMs can get better zero-shot abilities by instruction tuning. In the subsequent sections, we will delve into representative works of these three paradigms in detail.


\subsubsection{Fine-tuning}
Since under the fine-tuning paradigm, the utilization and training methodologies of generative LLMs are fundamentally similar to the discriminative LLMs discussed in Section~\ref{sec:Discriminative Fine-tuning}~\cite{zhang2024notellm}, therefore, we will only introduce a few representative studies in this subsection. For example, \cite{petrov2023generative} proposed GPTRec, which is a generative sequential recommendation model based on GPT-2. In contrast with BERT4Rec, which is based on discriminative LLM, GPTRec is based on generative LLM, uses SVD Tokenisation for memory efficiency, and is more flexible using the Next-K generation strategy. \cite{DBLP:journals/corr/abs-2305-06474} proposed to format the user historical interactions as prompts, where each interaction is represented by information about the item, and formulated the rating prediction task as two different tasks, respectively multi-class classification and regression. \cite{DBLP:journals/corr/abs-2305-06474} further investigated various LLMs in different sizes, ranging from 250M to 540B parameters and
evaluate their performance in zero-shot, few-shot, and fine-tuning scenarios, and found that the FLAN-T5-XXL (11B) model with fine-tuning can achieve the best result. 
\cite{li2023exploring} studied the influence of LLMs, such as GPT-3 with 175-billion parameters, on text-based collaborative filtering (TCF). \cite{DBLP:journals/corr/abs-2305-06566} proposed initially using closed-source LLMs such as ChatGPT to supplement item information in recommendation systems, obtaining closed-source tokens. subsequently, open-source LLMs like LLaMA were used for representing items and users, resulting in open-source embeddings. Finally, \cite{DBLP:journals/corr/abs-2305-06566} employed the fine-tuning paradigm to train the recommendation model based on the user and item embeddings. \cite{li2023exploring} found that using more powerful LLMs as text encoders can result in higher recommendation accuracy. However, an extremely
large LM may not result in a universal representation of users and items, and the simple ID-based collaborative filtering still remains a highly competitive approach in the warm item recommendation setting.

\subsubsection{Prompt Tuning}
In this paradigm, LLMs typically take the user/item information as input, and output the user preference (e.g., like or unlike, ratings) for the items, or output items that the user may be interested in. For example, \cite{DBLP:journals/corr/abs-2305-00447} proposed TALLRec which is trained by two tuning stages. TALLRec is first fine-tuned based on the self-instruct data by Alpaca~\cite{alpaca}. Then, TALLRec is further fine-tuned by recommendation tuning, where the input is the historical sequence of users and the output is the ``yes or no" feedback. 
\cite{ji2023genrec} presented an LLM-based generative recommendation method named GenRec that utilized the generation ability of generative LLM to directly generate the target item to recommend. Specifically, \cite{ji2023genrec} proposed to use input generation function to convert items into prompts, and use LLMs to generate the next item. 
\cite{chen2023palr} proposed a multi-step
approach to harness the potential of LLMs for recommendation. Specifically, \cite{chen2023palr} first proposed to leverage LLMs to generate a summary of a user’s preferences. For example, by analyzing a user’s music and TV viewing history, the LLM can generate a summary like ``pop music" and ``fantasy movies". Then, a retrieval module is utilized to get a much smaller candidate pool. Finally, the interaction history, natural language user profile, and retrieved candidates are utilized to construct a natural language prompt that can be fed into the LLM for recommendation. 
Similarly, drawing inspiration from the successful application of Convolutional Neural Networks (CNN) and Recurrent Neural Networks (RNN) models in user modeling, \cite{zheng2024harnessing} proposed two unique techniques for user preference summarization, respectively hierarchical summarization and recurrent summarization. \cite{zheng2024harnessing} further utilized SFT techniques to finetune the final recommendation model. \cite{chu2023leveraging} proposed to combine the user features and behavioral sequences as the input text. \cite{chu2023leveraging} also proposed to name the attributes in user features and items in the behavioral sequences as entities, and keep the entities as complete units in the input text. \cite{jin2023amazon} proposed to generate the title of the next product of interest for the user with the help of LLMs. They fine-tune a mT5 model using a generative objective defined on their dataset. However, a simple heuristic method that takes the last product title as a result surpasses the performance of the fine-tuned language model. 
\cite{friedman2023leveraging} proposed RecLLM, which contains a dialogue management module that uses an LLM to converse with the user, a ranker module that uses an LLM to match the user preferences, and a controllable LLM-based user simulator to generate synthetic conversations for tuning system modules. 

Further, \cite{li2023pbnr} proposed PBNR, which can describe user behaviors and news textually in the designed prompts. Specifically, the personalized prompts are created by designing input-target templates, wherein the relevant fields in the prompts are replaced with corresponding information from the raw data. To enhance the performance of LLMs on the recommendation task, PBNR incorporates the ranking loss and the language generation loss throughout the training. 
\cite{li2023gpt4rec} proposed to regard the recommendation task as a query generation and searching problem. They further utilized the LLMs to produce diverse and interpretable user interests representations, i.e., the queries. \cite{yue2023llamarec} argued that despite using instructions to prompt LLMs, the generated output does not directly provide ranking scores for candidates. In order to get the ranking scores of different items efficiently, \cite{yue2023llamarec} utilized the output from the LLM head, i.e., output scores over all items, as the ranking scores for candidates. \cite{li2024large} focused on using large language models for Point-of-Interest (POI) recommendation tasks.  The proposed framework constructed prompts to retain the heterogeneity of Location-based Social Networks (LBSN) data, avoiding the loss of contextual information and enabling an understanding of the intrinsic meaning of context. Additionally, by utilizing the concept of prompt-based trajectory similarity, it integrates historical trajectories with different users' trajectory information, thus alleviating the cold start problem and improving the prediction accuracy for trajectories of various lengths. 


In addition to directly fine-tuning the LLMs, some studies also proposed to utilize prompt learning to achieve better performance. For example, \cite{DBLP:conf/kdd/WangZWZ22} designed a unified conversational recommendation system named UniCRS based on knowledge-enhanced prompt learning. In this paper, the authors proposed to freeze the parameters of LLMs, and train the soft prompts for response generation and item recommendation by prompt learning. \cite{li2023personalized} proposed to provide user-understandable explanations based on the generation ability of LLMs. The authors tried both discrete prompt learning and continuous prompt learning, and further proposed two training strategies, respectively sequential tuning and recommendation as regularization.

Another noteworthy point is that how to control the output of large language models remains an unresolved challenge. When LLMs are used to directly generate items that users might be interested in, they may produce items that are out-of-corpus. To address this difficulty, some researchers have focused on how to combine prompting tuning with grounding methods, so that the results generated by the LLMs can be precisely aligned with the items in the item database. For example, \cite{lin2023multi} pointed out that item indexing and generation grounding are two essential steps for bridging LLMs and recommendation models. \cite{lin2023multi} first designed a multi-facet item indexing paradigm, which contains numeric ID, item title, and item attribute. Then, \cite{lin2023multi} pointed out that out-of-corpus identifiers and over-reliance on the quality of initially generated tokens are two critical problems in the generation process. To solve these problems, \cite{lin2023multi} proposed an FM-index-based multi-facet grounding method which can solve the above two problems simultaneously. \cite{bao2023bi} proposed a bi-step grounding paradigm. Specifically, \cite{bao2023bi} first proposed empowering LLMs to generate meaningful tokens through prompt tuning. Then, the output of LLMs was aligned with the real-world items by calculating the L2 distance between their embeddings, and some statistical information like the popularity factor was also utilized to rewight the L2 distance.

Furthermore, some studies have focused on integrating traditional collaborative models,  especially ID-based recommendation models, with LLM-based recommendation models by prompt tuning. For instance, \cite{zhang2023collm,zhang2023bridging} proposed to expand the vocabulary of LLMs to include user and item IDs. Subsequently, the embedding of these new tokens is trained through multi-step or joint training methods with both traditional and LLM-based recommendation models.
\cite{zhu2023collaborative} proposed to use both soft and hard prompting strategies to effectively learn user/item collaborative/content token embedding via language modeling on RS-specific corpora.
\cite{liao2023llara} proposed a hybrid item representation method, which integrates both textual tokens and behavioral tokens derived from the ID-based item embedding learned by traditional recommender models. To align the ID representation with the LLM token space, \cite{liao2023llara} designed an adapter based on a trainable linear projector. \cite{liao2023llara} further designed a curriculum prompt tuning, i.e., gradually shifting the learning focus from text-only prompting to hybrid prompting for better performance. Similarly, \cite{li2023e4srec} extracted the ID embeddings from a pre-trained sequential recommendation model, and employed a linear projection to convert the ID embeddings into the same dimension with the LLM token space. For the prediction, \cite{li2023e4srec} employed an item linear projection to replace the original prediction layer in LLMs via a weight matrix to get the ranking score of all the candidate items.
LLMGR~\cite{DBLP:journals/corr/abs-2402-16539} proposed integrating ID and graph embeddings with LLMs to leverage the complementary strengths of LLMs in natural language understanding and GNNs in relational data processing. \cite{qu2024elephant} founded that existing sequence recommendation methods based on pre-trained language models have not yet fully utilized the capabilities of language models and suffer from parameter redundancy. Based on this finding, the authors suggest using behavior-tuned PLMs (behavior-tuned pre-trained language models) to initialize item embeddings, thereby enhancing the capabilities of traditional sequence recommendation models such as SASRec, improving performance without increasing additional inference costs.

Lastly, we propose that most of the aforementioned methods are recommendations for general tasks using large language models. However, as previously mentioned, a significant advantage of large language models is their ability to efficiently align model parameters with specific domains, and some works mainly focus on the application of LLMs in specific domains. Take online recruitment as an example, within the realm of job-resume matching, the generative recommendation model GIRL~\cite{zheng2023generative} pioneers the use of LLM to generate potential job descriptions (JDs), enhancing the explainability and appropriateness of recommendations. GLRec~\cite{wu2023exploring} introduced the meta-path prompt constructor, a novel approach that employed LLM recommenders to interpret behavior graphs. This method also incorporated a path augmentation module to mitigate prompt bias. Subsequently, an LLM-based framework was introduced to align unpaired low-quality resumes with high-quality generated ones using Generative Adversarial Networks (GANs). This alignment process refined resume representations, leading to improved recommendation outcomes~\cite{du2023enhancing}.

\subsubsection{Instruction Tuning}
In this paradigm, LLMs are fine-tuned for multiple tasks with different types of instructions. In this way, LLMs can better align with human intent and achieve better zero-shot ability. For example, \cite{DBLP:conf/recsys/Geng0FGZ22} proposed to fine-tune a T5 model on five different types of instructions, respectively sequential recommendation, rating prediction, explanation generation, review summarization, and direct recommendation. After the multitask instruction tuning on recommendation datasets, the model can achieve the capability of zero-shot generalization to unseen personalized prompts and new items. Similarly, \cite{DBLP:journals/corr/abs-2205-08084} proposed to fine-tune an M6 model on three types of tasks, respectively scoring tasks, generation tasks, and retrieval tasks. \cite{DBLP:journals/corr/abs-2305-07001} first designed a general instruction format from three types of key aspects, respectively preference, intention, and task form. Then, \cite{DBLP:journals/corr/abs-2305-07001} manually designed 39 instruction templates and automatically generated a large amount of user-personalized instruction data for instruction tuning on a 3B FLAN-T5-XL model. The experiment results demonstrated that this approach can outperform several competitive baselines including GPT-3.5. \cite{yin2023heterogeneous} proposed to extract heterogeneous knowledge from the Meituan dataset, and constructed behavior text by prompt engineering. Then, \cite{yin2023heterogeneous} utilized instruction tuning to make the LLMs more effective for tasks in recommendation scenarios. \cite{li2023prompt} proposed to distill the discrete prompt for a specific task to a set of continuous prompt vectors so as to bridge IDs and words, and \cite{li2023prompt} leveraged instruction tuning to solve three different recommendation tasks, respectively sequential recommendation, top-N recommendation, and explanation generation. \cite{lu2024aligning} proposed that previous work has mainly focused on improving the accuracy of LLM-based recommendations without adequately addressing their instruction-following capabilities. Therefore, this paper introduces a reinforcement learning (RL) training strategy to significantly enhance the instruction-following capabilities of LLMs while performing recommendation tasks.

\linespread{1.22}
\begin{table*}[!ht]
\footnotesize
    \centering
    \caption{A list of representative LLM-based recommendation methods and their features. Note that, here the target of tuning/non-tuning denotes the used LLM module in the following methods.}
    \begin{tabular}{m{1.cm}<{\centering}|m{.7cm}|m{2.5cm}|m{3.9cm}|m{2.7cm}<{\centering}}
        \hline
        Adaption  &  \multicolumn{1}{c|}{Paper}  & \multicolumn{1}{c|}{Base Model} &  \multicolumn{1}{c|}{Recommendation Task} & Modeling Paradigm \\
        \hline
         \multicolumn{5}{c}{Discriminative LLMs for Recommendation}\\ \hline
        \multirow{6}{*}{\tabincell{c}{Fine-\\tuning}} & \cite{wu2021empowering} & BERT/UniLM & News Recommendation & LLM Embeddings + RS  \\ 
        \cline{2-5} & \cite{qiu2021u} & BERT  & User Representation & LLM Embeddings + RS  \\ 
        \cline{2-5} & \cite{zhang2022gbert} & BERT & Group Recommendation & LLM as RS  \\ 
        \cline{2-5} & \cite{yao2022reprbert} & BERT & Search/Matching & LLM Embeddings + RS  \\
        \cline{2-5} & \cite{muhamed2021ctr} & BERT & CTR Prediction & LLM Embeddings + RS  \\ 
        \cline{2-5} & \cite{xiao2022training} & BERT/RoBERTa & Conversational RS & LLM Embeddings + RS \\ 
        \hline
        \multirow{3}{*}{\tabincell{c}{Prompt\\ Tuning}} & \cite{zhang2023prompt} & BERT & Sequential Recommendation & LLM as RS \\ 
         \cline{2-5} & \cite{yang2022improving} & DistilBERT/GPT-2 & Conversational RS & LLM as RS \\
        \cline{2-5} & \cite{shen2023towards} & BERT & Conversational RS & LLM Embeddings + RS \\
         \cline{2-5} & \cite{penha2020does} & BERT & Conversational RS & LLM as RS \\ 
         \hline
         
         \multicolumn{5}{c}{Generative LLMs for Recommendation}\\ \hline
          \multirow{12}{*}{\tabincell{c}{Non-\\tuning}} & \cite{DBLP:journals/corr/abs-2305-06566} & \tabincell{l}{ChatGPT} & News Recommendation & LLM Tokens + RS  \\ 
        \cline{2-5} & \cite{DBLP:conf/kdd/WangZWZ22} & \tabincell{l}{DialoGPT/RoBERTa}  & \tabincell{l}{Converational RS} & LLM Tokens + RS / LLM as RS \\  
        \cline{2-5} & \cite{DBLP:conf/ecir/SileoVR22} & \tabincell{l}{GPT-2}  & \tabincell{l}{Sequential Recommendation} & LLM as RS \\  
        \cline{2-5} & \cite{DBLP:journals/corr/abs-2304-03153} & \tabincell{l}{GPT-3.5}  & \tabincell{l}{Sequential Recommendation} & LLM Tokens + RS / LLM as RS \\  
        \cline{2-5} & \cite{DBLP:journals/corr/abs-2303-14524} & \tabincell{l}{ChatGPT/GPT-3.5}  & \tabincell{l}{Sequential Recommendation} & LLM as RS \\  
        \cline{2-5} & \cite{DBLP:journals/corr/abs-2304-03516} & \tabincell{l}{ChatGPT}  & \tabincell{l}{Generative Recommendation} & LLM as RS \\  
        \cline{2-5} & \cite{DBLP:journals/corr/abs-2305-08845} & \tabincell{l}{ChatGPT}  & \tabincell{l}{Sequential Recommendation} & LLM as RS \\  
        \cline{2-5} & \cite{DBLP:journals/corr/abs-2304-09542} & \tabincell{l}{ChatGPT/GPT-3.5}  & \tabincell{l}{Passage Reranking} & LLM as RS \\  
        \cline{2-5} & \cite{qin2023large} & \tabincell{l}{T5/GPT-3.5/GPT-4}  & \tabincell{l}{Passage Reranking} & LLM as RS \\ 
        \cline{2-5} & \cite{DBLP:journals/corr/abs-2304-10149} & \tabincell{l}{ChatGPT}  & \tabincell{l}{Five Tasks} & LLM as RS \\  
        \cline{2-5} & \cite{DBLP:journals/corr/abs-2305-02182} & \tabincell{l}{ChatGPT/GPT-3.5}  & \tabincell{l}{Sequential Recommendation} & LLM as RS \\ 
        \cline{2-5} & \cite{Xi2023TowardsOR} & \tabincell{l}{ChatGLM}  & \tabincell{l}{CTR Prediction} & LLM Tokens + RS \\ 
        \cline{2-5} & \cite{wang2023recagent} & \tabincell{l}{ChatGPT}  & \tabincell{l}{Recommendation Agent} & LLM Tokens + RS \\ 
        \cline{2-5} & \cite{DBLP:journals/corr/abs-2310-10108} & \tabincell{l}{ChatGPT}  & \tabincell{l}{Recommendation Agent} & LLM Tokens + RS \\ 
        \hline
         
       \multirow{10}{*}{\tabincell{c}{Tuning}} & \cite{DBLP:journals/corr/abs-2305-07001} &  \tabincell{l}{FLAN-T5} & Three Tasks & LLM as RS  \\ 
        \cline{2-5} & \cite{DBLP:journals/corr/abs-2305-06474} & \tabincell{l}{FLAN-T5/ChatGPT} & Rating Prediction &  LLM as RS \\ 
        \cline{2-5} & \cite{DBLP:journals/corr/abs-2305-00447} & \tabincell{l}{LLaMA-7B} & Movie/Book RS &  LLM as RS \\
        \cline{2-5} & \cite{li2023personalized} & \tabincell{l}{GPT-2} &Explainable RS  &  LLM as RS \\
        \cline{2-5} & \cite{DBLP:conf/recsys/Geng0FGZ22} & \tabincell{l}{T5} & \tabincell{l}{Five Tasks} &  LLM as RS \\
        \cline{2-5} & \cite{DBLP:journals/corr/abs-2205-08084} & \tabincell{l}{M6} & \tabincell{l}{Five Tasks} &  LLM as RS \\
        \cline{2-5} & \cite{lin2023multi} & \tabincell{l}{BART/LLaMA} & \tabincell{l}{Text-based/Sequential Recommendation} &  LLM as RS \\
        \cline{2-5} & \cite{wu2023exploring} & \tabincell{l}{BELLE} & \tabincell{l}{Job Recommendation} &  LLM as RS \\
        \cline{2-5} & \cite{zheng2023generative} & \tabincell{l}{BELLE} & \tabincell{l}{Generative Recommendation} &  LLM Tokens +RS \\
        \cline{2-5} & \cite{Mao2023UniTRecAU} & \tabincell{l}{UniTRecAU} & \tabincell{l}{Text-based Recommendation} &  LLM as RS \\
        \cline{2-5} & \cite{bao2023bi} & \tabincell{l}{LLaMA-7B} & Movie/Game RS &  LLM as RS \\
        \cline{2-5} & \cite{li2023exploring} & \tabincell{l}{OPT} & \tabincell{l}{Text-based Recommendation} &  LLM Embeddings +RS \\
        \cline{2-5} & \cite{li2023prompt} & \tabincell{l}{T5-small} & \tabincell{l}{Three Tasks} &  LLM as RS \\
        \cline{2-5} & \cite{li2023ctrl} & \tabincell{l}{RoBERTa/GLM} & \tabincell{l}{CTR Prediction} &  LLM Embeddings +RS \\
        \hline
    \end{tabular}
    
    \label{tab:list}
\end{table*}

\section{Findings}
\begin{figure*}[t]
    \centering
    \includegraphics[width=1.\textwidth]{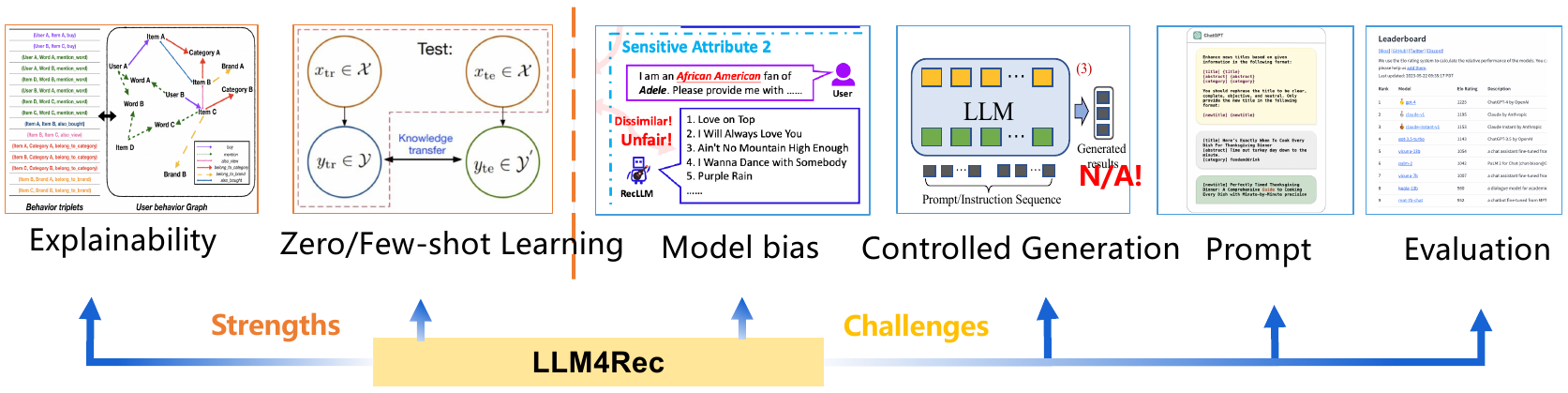}
    \caption{The major strengths and technical challenges of LLM4Rec.}
    \label{fig:findings}
\end{figure*}

In this survey, we systematically reviewed the application paradigms and adaptation strategies of large language models in recommendation systems, especially for generative language models. We have identified their potential to improve the performance of traditional recommendation models in specific tasks. However, it is necessary to note that the overall exploration in this field is still in the early stage. Researchers may find it challenging to determine the most worthwhile problems and pain points to investigate. To address this, we have summarized the common findings presented by numerous studies on large-scale model recommendations. As shown in Figure~\ref{fig:findings}, these findings highlight certain technical challenges and present potential opportunities for further advancements in the field.

\linespread{1.05}
\begin{table*}[!ht]
\footnotesize
    \centering
    \caption{A list of common datasets used in existing LLM-based recommendation methods.}
    \begin{tabular}{m{1.1cm}<{\centering}|m{1.15cm}|m{1.15cm}|m{4.9cm}|m{2.2cm}}
        \hline
        Name  &  \multicolumn{1}{c|}{Scene}  & \multicolumn{1}{c|}{Tasks} &  \multicolumn{1}{c|}{Information} & URL \\
        \hline
        Amazon Review \cite{he2016ups} & Commerce  & Seq Rec / CF Rec & This is a large crawl of product reviews from Amazon. Ratings: 82.83 million, Users:	20.98 million, Items: 9.35 million, Timespan: May 1996 - July 2014 &  \url{http://jmcauley.ucsd.edu/data/amazon/}  \\ 
        \hline
        Amazon-M2~\cite{jin2023amazon} & Commerce  & Seq Rec / CF Rec & A large dataset of anonymized user sessions with their interacted products collected from multiple language sources at Amazon. It includes 3,606,249 train sessions, 361,659 test sessions, and 1,410,675 products. &  \url{https://arxiv.org/abs/2307.09688}  \\ 
        \hline
        Amazon Review 2023~\cite{DBLP:journals/corr/abs-2403-03952} & Commerce  & Seq Rec / CF Rec & The dataset comprises over 570 million reviews and 48 million items from 33 categories. & \href{https://amazon-reviews-2023.github.io}{https://amazon-reviews -2023.github.io}  \\ 
        \hline
        Steam \cite{wan2018item} & Game  & Seq Rec / CF Rec & Reviews represent a great opportunity to break down the satisfaction and dissatisfaction factors around games. Reviews: 7,793,069, Users:	2,567,538, Items: 15,474, Bundles: 615 &  \url{https://cseweb.ucsd.edu/~jmcauley/datasets.html#steam_data}  \\ 
        \hline
        MovieLens & Movie  & General & The dataset consists of 4 sub-datasets, which describe users' ratings to moives and free-text tagging activities from MovieLens, a movie recommendation service.  &  \url{https://grouplens.org/datasets/movielens/}  \\ 
        \hline
        Yelp & Commerce  & General & There are 6,990,280 reviews, 150,346 businesses, 200,100 pictures, 11 metropolitan areas, 908,915 tips by 1,987,897 users. 
Over 1.2 million business attributes like hours, parking, availability, etc. &  \url{https://www.yelp.com/dataset}  \\ 
        \hline
        Douban \cite{wu2017sequential} & Movie, Music, Book  & Seq Rec / CF Rec & This dataset includes three domains, i.e., movie, music, and book, and different kinds of raw information, i.e., ratings, reviews, item details, user profiles, tags (labels), and date. &  \href{https://github.com/MarkWuNLP/MultiTurnResponseSelection}{https://github.com/ MarkWuNLP/ MultiTurnResponse Selection}  \\ 
        \hline
        MIND \cite{wu2020mind} & News  & General & MIND contains about 160k English news articles and more than 15 million impression logs generated by 1 million users. Every news contains textual content including title, abstract, body, category, and entities. &  \url{https://msnews.github.io/assets/doc/ACL2020_MIND.pdf}  \\
        \hline
        U-NEED \cite{liu2023u} & Commerce  & Conversa -tion Rec & U-NEED consists of 7,698 fine-grained annotated pre-sales dialogues, 333,879 user behaviors, and 332,148 product knowledge tuples. &  \url{https://github.com/LeeeeoLiu/U-NEED}  \\
        \hline
        KuaiSAR \cite{Sun2023KuaiSAR} & Video  & Search and Rec & KuaiSAR contains genuine search and recommendation behaviors of 25,877 users, 6,890,707 items, 453,667 queries, and 19,664,885 actions within a span of 19 days on the Kuaishou app. &  \url{https://kuaisar.github.io/}  \\
        \hline
        Tenrec \cite{yuan2022tenrec} & Video, Article  & General & Tenrec is a large-scale benchmark dataset for recommendation systems. It contains around 5 million users and 140 million interactions. &  \url{https://tenrec0.github.io/}  \\
        \hline
        PixelRec \cite{cheng2023image} & Video  & Seq Rec / CF Rec & PixelRec is a massive image-centric recommendation dataset that includes approximately 200 million user-image interactions, 30 million users, and 400,000 cover images.  The texts and other aggregated attributes of videos are also included. &  \url{https://github.com/westlake-repl/PixelRec}  \\
        \hline
    \end{tabular}
    
    \label{tab:data}
\end{table*}
\subsection{Model Bias}
\noindent\textbf{Position Bias.} In the generative language modeling paradigm of recommendation systems, various information such as user behavior sequences and recommended candidates are input to the language model in the form of textual sequential descriptions~\cite{harte2023leveraging}, which can introduce some position biases inherent in the language model itself~\cite{lu2022fantastically}. For example, the order of candidates affects the ranking results of LLM-based recommendation models, i.e., LLM often prioritizes the items in the top order. And the model usually cannot capture the behavior order of the sequence well.~\cite{hou2022towards} used the random sampling-based bootstrapping to alleviate the position bias of candidates and emphasized the recently interacted items to enhance behavior order. However, these solutions are not adaptive enough to adapt to different contexts, and more robust learning strategies are needed in the future.

\noindent\textbf{Popularity Bias.} The ranking results of LLMs are influenced by the popularity levels of the candidates. Popular items, which are often extensively discussed and mentioned in the pre-training corpora of LLMs, tend to be ranked higher. This can lead to a lack of diversity in the responses and potentially marginalize less popular or minority viewpoints. Addressing this issue is challenging as it is closely tied to the composition of the pre-trained corpus.

\noindent\textbf{Fairness Bias.}
Pre-trained language models have exhibited fairness issues related to sensitive attributes~\cite{DBLP:journals/corr/abs-2305-07609,DBLP:journals/corr/abs-2403-05668}, which are influenced by the training data or the demographics of the individuals involved in certain task annotations~\cite{ferrara2023should}. These fairness concerns can result in models making recommendations that assume users belong to a specific group, potentially leading to controversial issues when deployed commercially. One example is the bias in recommendation results caused by gender or race~\cite{shen2023towards}. Addressing these fairness issues is crucial and necessary to ensure equitable and unbiased recommendations.

\noindent\textbf{Personalization Bias.}
Introducing collaborative filtering signals into large language models (LLMs) for recommendation purposes presents several challenges, particularly when compared to traditional ID-based recommendation models. While LLMs have the potential to generate highly personalized content by understanding nuanced textual inputs, translating this capability into personalized recommendations is challenging. Traditional models, with their direct mapping of user-item interactions, can sometimes more straightforwardly personalize recommendations. Integrating with ID-based recommendation models or directly training and learning ID-based tokens are potential approaches in this field~\cite{li2023exploring}.

\subsection{Recommendation Prompt Designing}
\noindent\textbf{User/Item Representation.}
In practice, recommendation systems typically utilize a large number of discrete and continuous features to represent users and items. However, most existing LLM-based work only uses the name to represent items, and a list of item names to represent users, which is insufficient for modeling users and items accurately. 
Additionally, it is critical to translate a user's heterogeneous behavior sequence (such as clicks, adding to cart, and purchases in the e-commerce domain) into natural language for preference modeling. 
ID-like features have been proven effective in traditional recommendation models, but incorporating them into prompts to improve personalized recommendation performance is also challenging.

\noindent\textbf{Limited Context Length.}
The context length limitation of LLMs will constrain the length of users' behavioral sequences and the number of candidate items, resulting in suboptimal performance~\cite{DBLP:journals/corr/abs-2305-07001}.
Existing work has proposed some techniques to alleviate this problem, such as selecting representative items from user behavior sequence~\cite{DBLP:journals/corr/abs-2304-03153} and sliding window strategy for candidate list~\cite{DBLP:journals/corr/abs-2304-09542}.
Recently, there have been efforts to extend the context length limitations of LLMs. For example, the LongLLaMA model~\cite{tworkowski2023focused} is a large language model capable of handling long contexts of 256k tokens or even more, which is fine-tuned using the Focused Transformer (FoT) method.
However, the effectiveness of these methods in application to recommendation scenarios still merits further validation and study.

\subsection{Promising Ability}
\noindent\textbf{Zero/Few-shot Recommendation Ability.}
The experimental results on multiple domain datasets indicate that LLMs possess impressive zero/few-shot abilities in various recommendation tasks~\cite{DBLP:journals/corr/abs-2305-08845,DBLP:journals/corr/abs-2304-10149,DBLP:journals/corr/abs-2305-02182}.
It is worth noting that few-shot learning, which is equivalent to in-context learning, does not change the parameters of LLMs.
This suggests LLMs have the potential to mitigate the cold-start problem with limited data.
However, there are still some open questions, such as the need for clearer guidance in selecting representative and effective demonstration examples for few-shot learning, as well as the need for experimental results across more domains to further support the conclusion regarding the zero/few-shot recommendation abilities.


\noindent\textbf{Explainable Ability.}
Generative LLMs exhibit a remarkable ability for natural language generation.
Thus, a natural thought is to use LLMs to conduct explainable recommendations via a text-generation manner~\cite{silva2024leveraging,wang2024can}.
~\cite{DBLP:journals/corr/abs-2304-10149} conduct a comparison experiment among ChatGPT and some baselines on explanation generation task.
The results demonstrate that even without fine-tuning and under the in-context learning setting, ChatGPT still performs better than some traditional supervised methods.
Moreover, according to human evaluation, ChatGPT's explanations are deemed even clearer and more reasonable than the ground truth.
Encouraged by these exciting preliminary explorations and  experimental results, the performance of fine-tuned LLMs in explainable recommendations is expected to be promising.

\subsection{Evaluation Issues}
\noindent\textbf{Generation Controlling.}
As we mentioned before, many studies have employed large-scale models as recommendation systems by providing carefully designed instructions. For these LLMs, the output should strictly adhere to the given instruction format, such as providing binary responses (yes or no) or generating a ranked list. However, in practical applications, the output of LLMs may deviate from the desired output format. For instance, the model may produce responses in incorrect formats or even refuse to provide an answer~\cite{DBLP:journals/corr/abs-2305-02182}. And, generative models struggle to perform well in list-wise recommendation tasks due to their training data and autoregressive training mode, which make them less capable of handling ranking problems with multiple items. This issue cannot be resolved through fine-tuning, as there is no ground truth for ranking multiple items in a sequence in real-world scenarios. Therefore, it is difficult to apply autoregressive training logic based on sequence. PRP (Pairwise Ranking Prompting)~\cite{qin2023large} proposes pairwise ranking for listwise tasks with LLM, which enumerates all pairs and performs a global aggregation to generate a score for each item. However, this logic is time consuming in the inference process. Therefore, addressing the challenge of ensuring better control over the output of LLMs is a pressing issue that needs to be resolved.

\noindent\textbf{Evaluation Criteria.}
If the tasks performed by LLMs are standard recommendations, such as rating prediction or item ranking, we can employ existing evaluation metrics for evaluation, e.g., NDCG, MSE, etc. However, LLMs also have strong generative capabilities, making them suitable for generative recommendation tasks~\cite{DBLP:journals/corr/abs-2304-03516}. Following the generative recommendation paradigm, LLMs can generate items that have never appeared in the historical data and recommend them to users. In this scenario, evaluating the generative recommendation capability of LLMs remains an open question.

\noindent\textbf{Datasets.}
Currently, most of the research in this area primarily tests the recommendation capability and zero/few-shot capability of LLMs using datasets like MovieLens, Amazon Books, and similar benchmarks. However, this may bring the following two potential issues. First, compared to real-world industrial datasets, these datasets are relatively small in scale and may not fully reflect the recommendation capability of LLMs. Second, the items in these datasets, such as movies and books, may have related information that appeared in the pre-training data of LLMs. This could introduce bias in evaluating the few-zero-shot learning capability of LLMs. Currently, we still lack a suitable benchmark for conducting a more comprehensive evaluation.


In addition to the aforementioned prominent findings, there are also some limitations associated with the capabilities of large language models. For example, the challenge of knowledge forgetting may arise when training models for specific domain tasks or updating model knowledge~\cite{jang2022towards}. Another issue is the distinct performances caused by varying sizes of language model parameters, where using excessively large models would result in excessive computational costs for research and deployment in recommendation systems~\cite{DBLP:journals/corr/abs-2305-08845}. These challenges also present valuable research opportunities in the field.

\section{Conclusion}

In this paper, we reviewed the research area of large language models (LLMs) for recommendation systems. We classified existing work into discriminative models and generative models, and then illustrated them in detail by the domain adaption manner. To prevent conceptual confusion, we provided definitions and distinctions of fine-tuning, prompting, prompt tuning, and instruction tuning in LLM-based recommendations. To the best of our knowledge, our survey is the first systematic and up-to-date review specifically dedicated to generative LLMs for recommendation systems, which further summarized the common findings and challenges presented by numerous related studies. Therefore, this survey provided researchers with a valuable resource for gaining a comprehensive understanding of LLM recommendations and exploring potential research directions.

Looking to the future, as computational capabilities continue to advance and the realm of artificial intelligence expands, we anticipate even more sophisticated applications of LLMs in recommendation systems. There is a promising horizon where the adaptability and precision of these models will be harnessed in more diverse domains, possibly leading to real-time, personalized recommendations that consider multi-modal inputs. Moreover, as ethical considerations gain prominence, future LLM-based recommendation systems might also integrate fairness, accountability, and transparency more intrinsically.

In conclusion, while we have made substantial strides in understanding and implementing LLMs in recommendation systems, the journey ahead is replete with opportunities for innovation and refinement. Our survey, we hope, will serve as a foundational stepping stone for the next wave of discoveries in this dynamic and ever-evolving field.








\bibliographystyle{sn-mathphys-num}
\bibliography{main}

\end{document}